\documentclass[superscriptaddress,
               floatfix,
               longbibliography, 
               showkeys,apl,twocolumn]{revtex4-2}

\usepackage{graphicx} 
\usepackage{bm}	
\usepackage{color}                     
\usepackage{xcolor}
\usepackage{epsfig}
\usepackage{amsmath} 
\usepackage{amssymb} 
\usepackage{longtable} 
\usepackage{float}
\usepackage{mathtools}
\usepackage{dsfont}
\usepackage{xparse}
\usepackage{hyperref}
\usepackage{times}
\usepackage[normalem]{ulem}
\usepackage{sidecap}
\usepackage{appendix} 
\sidecaptionvpos{figure}{t}
\usepackage{comment}
\usepackage{physics}
\usepackage{float}
\usepackage{xfrac}

\usepackage{multirow}
\usepackage{algpseudocode}
\usepackage{tikz,lipsum,lmodern}
\usepackage[most]{tcolorbox}
\usepackage{tikz}

\definecolor{darkgreen}{rgb}{0.0, 0.5, 0.0}

\newcommand{\be}{\begin{equation}}
\newcommand{\ee}{\end{equation}}
\newcommand{\bd}{\begin{displaymath}}
\newcommand{\ed}{\end{displaymath}}
\newcommand{\BE}{\begin{eqnarray}}
\newcommand{\EE}{\end{eqnarray}}

\usetikzlibrary{quantikz2}

\begin{document}
    \title{Universal Parent Hamiltonians for Adiabatic Warm Starts}
    \author{Feng Qian}
    \email{feng.qian@tufts.edu}
    \affiliation{Department of Physics and Astronomy, Tufts University, Medford MA
}%
    \author{Peter J. Love}
    \email{peterjlove@gmail.com}
\affiliation{Department of Physics, University of Toronto, Toronto, ON M5S 1A7, Canada}
\affiliation{Department of Computer Science, University of Toronto, Toronto, ON M5S 3H6, Canada}
\affiliation{QMatter, Inc., Office 109, 254 Chapman Rd, Suite 101-B, Newark, DE 19702, USA}

    \date{\today}

    \begin{abstract}
Computing the ground state properties of quantum systems is an important potential application of quantum computing. The success probability of quantum phase estimation approaches to ground state problems is proportional to the overlap of the input state with the ground state. Local ansatz state approaches suffer from the orthogonality catastrophe, whereas adiabatic state preparation (ASP) can prepare good approximations with a cost growing with the inverse square of the minimum spectral gap along the adiabatic path. When the initial and target Hamiltonians lie in quantum phases separated by a first order phase transition, the minimum spectral gap along the adiabatic path becomes exponentially small as a function of system size. We pursue a solution to this problem based on choosing an initial Hamiltonian for adiabatic state preparation whose ground state lies in the same quantum phase as the target ground state. We develop a protocol for universal adiabatic warm starts with universal parent Hamiltonians (UPHAWS) that can initialize ASP in any state whose preparation circuit is known. We use the Feynman--Kitaev clock Hamiltonian as a universal parent Hamiltonian, for preparation circuits with and without mid circuit measurement. We benchmark the framework on a $\mathbb{Z}_2$-symmetric matrix product state (MPS) family interpolating to a target GHZ Hamiltonian, and on the linear $H_6$ chain under symmetric bond stretching. For the $H_6$ system a bond-dimension-$4$ matrix product state warm-start increases the minimum gap on the adiabatic path by a factor of two relative to the Hartree-Fock initialization. To perform these classical benchmark simulations we develop a momentum-space truncation of the adiabatic Hamiltonian that may be of independent interest.
\end{abstract}

    \maketitle

    \section{Introduction}
\label{sec:introduction}

Simulation of quantum systems is a major potential application of future quantum computers. One of the central challenges in quantum simulation is the estimation of the ground-state energy. Classical approximation methods---Hartree--Fock \cite{szabo1996modern}, the density matrix renormalization group (DMRG) \cite{schollwock2005density,olivares2015ab}, density functional theory (DFT) \cite{kohn1965self}, and coupled-cluster theory \cite{vcivzek1966correlation}---exploit problem structure to approximate the ground state, but many problems remain out of reach. 

Ground state problem may be solved on quantum computers by quantum phase estimation (QPE)~\cite{kitaev1995quantum}. QPE projects an input state onto an eigenstate of the target Hamiltonian, producing a digital estimate of the corresponding eigenvalue. The success probability of QPE depends on the overlap between the initial input state and the true ground state, making the preparation of high-overlap initial states a central algorithmic bottleneck.

One approach is ansatz state preparation: a parameterized circuit prepares an approximation of the ground state, with parameter count limited by the cost of optimizing the ansatz. Common examples include Hartree--Fock states, linear combinations of a few Slater determinants~\cite{Fomichev_2024}, and states with compact local-correlation structure such as matrix product states (MPS)~\cite{schon2005sequential}~ and unitary coupled cluster ansatz~\cite{Anand_2022}. 
Simple ansatz with product structure can exhibit the orthogonality catastrophe~\cite{kohn1999nobel}, in which the overlap of the ansatz with the true ground state is the product of the overlaps of its subsystems, resulting in an exponentially decreasing overlap with system size~\cite{mcclean2014exploiting,Lee_2023}. Furthermore, for quantum phase estimation to outperform classical methods, there must exist and ansatz that can achieve a good overlap with the true ground state, without also enabling classical calculation of the energy to the desired accuracy. Whether this is possible has recently been called into question~\cite{Lee_2023}. 

Heuristic quantum algorithms for state preparation include open quantum system dynamics ~\cite{zhan2026rapid}, quantum imaginary time evolution~\cite{motta2020determining} and adiabatic state preparation (ASP)~\cite{farhi2000quantumcomputationadiabaticevolution,aharonov2003adiabaticquantumstategeneration}. ASP relies on a time-dependent Hamiltonian that slowly interpolates between a simple initial Hamiltonian (with a trivially preparable ground state) and a target problem Hamiltonian whose ground state encodes the solution. If the system evolves slowly enough, the adiabatic theorem guarantees that the wavefunction will track the instantaneous ground state, yielding a good approximation of the true target ground state at the end of the evolution~\cite{messiah2014quantum}. The required physical runtime for adiabatic evolution is bounded by $\mathcal{O}(\Delta^{-2})$, where $\Delta$ is the minimum spectral gap between the ground and first excited states along the adiabatic path. However, analytically or numerically determining this minimal gap is at least as hard as ground state energy estimation~\cite{ambainis2014physicalproblemsslightlydifficult}. 

Condensed matter physics gives a picture of success and failure modes for ASP in terms of quantum phases. Quantum phases are ground-state manifolds distinguished by symmetry-breaking order parameters or topological invariants, separated by quantum critical points where the spectral gap closes~\cite{sachdev1999quantum}. Traversing a second-order transition causes the minimum gap to close polynomially with system size, while crossing a first-order phase boundary results in a minimum gap that shrinks exponentially with system size~\cite{Amin_2009,Zurek_2005}. Efficient ASP, in this picture, relies on constructing adiabatic paths that begin and end in the same quantum phase, or transition between phases that are separated only by second order phase transitions. If the initial state lies in the same phase as the target ground state, no phase boundary is crossed and the required evolution time of simulating the time dependent adiabatic evolution can be efficiently simulated on a quantum computer, yielding an efficient state preparation algorithm ~\cite{Osborne_2007,Hastings_2005,Jansen_2007,Bachmann_2011}. 

An example of a chemical system undergoing a quantum phase transition is a hydrogen chain in which the interatomic spacing is uniformly varied. At large spacing, the system resides in a tight-binding, highly localized phase, whereas at small spacing, the wavefunctions overlap to form a delocalized metallic phase~\cite{PhysRevX.10.031058,Hachmann_2006}. ASP of (de)-localized states starting from (de)-localized states will not cross a quantum critical point and so will succeed. These preparations correspond to adiabatic change in the interatomic spacing without changing the character of the ground state wavefunction. Attempting to adiabatically reach the delocalized (metallic) phase from the localized (tight-binding) phase will result in traversing a quantum phase transition and an inefficient ASP. This picture suggests that the initial state for ASP requires careful selection.

Numerical studies of Fe-S clusters in Nitrogenase in~\cite{Lee_2023} gave numerical evidence that the minimum spectral gap of a linear interpolation from an initial Hartree-Fock Hamiltonian $H_{\text{init}}$ to the molecular Hamiltonian $H_{\text{target}}$ scales polynomially with the overlap between the initial ground state $|\Upsilon_0\rangle$ and final ground state $|\Psi_0\rangle$. Specifically~\cite{Lee_2023} argue that $1/\Delta_{\rm min}\sim {\rm poly}\left( 1/|\langle\Upsilon_0 | \Psi_0\rangle\right)$ so that the time required for ASP is $T_{\text{ASP}} \sim {\rm poly}\left( 1/|\langle\Upsilon_0 | \Psi_0\rangle\right)$ (see~\cite{Lee_2023}, Figure 2). For the single-reference and configuration state function initial states used in~\cite{Lee_2023} numerical evidence supports the exponential decay of $\langle\Upsilon_0 | \Psi_0\rangle$ with system size as expected for systems exhibiting an orthogonality catastrophe.

The relationship between initial overlap and adiabatic fidelity was investigated for a broader class of systems in~\cite{PhysRevLett.119.200401,chen2022bounds}. Those authors consider systems that suffer from a generalized orthogonality catastrophe (GOC). A system undergoing adiabatic evolution exhibits a GOC if  the overlap between ground states on the adiabatic path decays exponentially with their separation (see~\cite{PhysRevLett.119.200401}, eq. $4$.). The overlap of the ground state along the path with the initial state $C(t)=|\langle\Upsilon_0 | \Psi_t\rangle|^2$ characterizes how ASP changes the initial state. The adiabatic fidelity $F(t)=|\langle\Phi_t | \Psi_t\rangle|^2$ is the overlap of the ground state along the path $| \Psi_t\rangle$ with the state of the system determined by time dependent evolution $| \Phi_t\rangle$.  The adiabatic fidelity characterizes how adiabatic the time-dependent evolution is, i.e. how well the adiabatic theorem is being obeyed. In~\cite{PhysRevLett.119.200401,chen2022bounds} the difference $|F(t)-C(t)|$ was bounded above by a quantum speed limit~\cite{PhysRevLett.70.3365,RevModPhys.67.759} for systems with a GOC. The results of~\cite{PhysRevLett.119.200401,chen2022bounds} imply that a generalized orthogonality catastrophe will cause ASP to fail because the adiabatic fidelity $F(t)$ must track $C(t)$. Hence if $C(t)$ is small, as it must be for ASP to generate states different from the initial state, and $|F(t)-C(t)|$ is tightly bounded above, then the evolution cannot be adiabatic.

To warm start ASP in a chosen phase, we need two things. Firstly, we need an ansatz for a state in the correct phase. Secondly, we need a so-called "parent Hamiltonian" for this state, i.e. a starting point for the adiabatic evolution. In quantum many-body physics, parent Hamiltonians are used to characterize entanglement and correlations: the Affleck--Kennedy--Lieb--Tasaki (AKLT) state~\cite{affleck1987rigorous} and other tensor network states~\cite{perezgarcia2007pepsuniquegroundstates,perez2006matrix} are unique ground states of frustration-free parent Hamiltonians. Previous work has constructed local parent Hamiltonians for specific classes of states---stabilizer states~\cite{Ni_2015}, finitely correlated states~\cite{Qi_2019}, and projected entangled-pair states (PEPS)~\cite{perezgarcia2007pepsuniquegroundstates}---by identifying local projectors that annihilate the target state. 

A central contention of~\cite{Lee_2023} is that it is difficult to identify sufficiently good initial states for ASP from classical ansatz without those ansatz states also enabling classical solution of the energy estimation problem. However, in quantum simulation there is no reason to restrict the initial states to those that can be efficiently classically constructed. One only requires that the state has an efficient quantum preparation circuit. This motivates the development of {\em universal} parent Hamiltonians. It is this question we shall tackle in the remainder of the paper.

The paper is organized as follows. Section~\ref{sec:background} reviews adiabatic state preparation, matrix product state preparation and the Feynman--Kitaev clock Hamiltonian. Section~\ref{sec:Theory} then develops the clock Hamiltonian as a parent Hamiltonian for arbitrary circuit-preparable states, analyzes its spectral gap for deterministic and probabilistic preparation circuits, and constructs the rotating-frame block matrix elements used for numerical simulation. Section~\ref{sec:numerical_results} presents numerical results: we study the adiabatic gap of a $\mathbb{Z}_2$-symmetric matrix-product-state family interpolating toward a target GHZ Hamiltonian, benchmark the momentum-space truncation scheme, and apply the framework to warm-started ground-state preparation of the linear $H_6$ chain under symmetric bond stretching. Section~\ref{sec:conclusion} summarizes the results and identifies some directions for future work.

    \section{Background}
\label{sec:background}


In this section we develop the theoretical ingredients used throughout the paper: adiabatic state preparation, the Feynman--Kitaev clock Hamiltonian, its use as a parent Hamiltonian for a circuit-prepared warm-start, the resulting spectral gap for deterministic and probabilistic preparation circuits, and the rotating-frame representation used in the numerical simulations of Sec.~\ref{sec:numerical_results}.

\subsection{Adiabatic State Preparation}
\label{sec:ASPPrep}

Adiabatic evolution is described by the time-dependent Schr\"odinger equation:
\begin{equation}
    i \frac{1}{\mathcal{T}} \frac{d}{ds}|\psi(s)\rangle = H_A(s)|\psi(s)\rangle,
\end{equation}
where $\mathcal{T}$ is the total physical evolution time and $s = t/\mathcal{T} \in [0,1]$ is a dimensionless parameter that controls the rate of change of the adiabatic Hamiltonian $H_A(s)$. A linear interpolation schedule between initial Hamiltonian $H_i$ and final Hamiltonian $H_f$ is defined as $H_A(s) = (1-s)H_{i} + sH_{f}$.

The adiabatic theorem states that if a system is initialized at $s = 0$ in the ground state of the initial Hamiltonian $H_{i}$, it will remain close to the instantaneous ground state of $H_A(s)$ throughout the evolution, provided the physical time $\mathcal{T}$ is sufficiently long and there is a non-vanishing spectral gap between the ground and first excited states along the path. At $s = 1$, the final state will be a good approximation to the target ground state of $H_{f}$. The degree of approximation can be measure by the adiabatic fidelity $F(s)=|\langle \psi(s)|\phi(s) \rangle$ where $\phi(s)$ is the ground state of $H(s)$.

Adiabatic State Preparation (ASP) is a quantum heuristic algorithm that uses this theorem to prepare the ground state of a problem Hamiltonian $H_p$ that encodes the solution of interest (setting $H_f = H_p$). Let us define the instantaneous eigenstates and eigenvalues of the adiabatic Hamiltonian as:
\begin{equation}
    H_A(s) |s; l\rangle = E(s; l)|s; l\rangle,
\end{equation}
where $l \in \{0, 1, \dots, \dim(\mathcal{H})-1\}$ labels the energy levels (e.g., $|s; 0\rangle$ is the instantaneous ground state). For successful ground state preparation, the required physical runtime $\mathcal{T}$ is bounded by:
\begin{equation}
    \mathcal{T} \sim \mathcal{O}\left( \frac{\epsilon}{\Delta^2} \right),
\end{equation}
where $\epsilon = \max_s \left| \langle s; 1 | \frac{dH_A(s)}{ds} | s; 0 \rangle \right| \sim \mathcal{O}(N)$ bounds the Hamiltonian's rate of change, and $\Delta = \min_s \big(E(s; 1) - E(s; 0)\big)$ is the minimum spectral gap. For a wide variety of problems, the required adiabatic runtime is determined by the inverse square of this minimum gap, $\Delta^{-2}$ \cite{farhi2000quantumcomputationadiabaticevolution, Albash_2018}. 

\subsection{Matrix Product State Preparation}
\label{sec:MPSPrep}

The goal of this work is to extend the initialization of ASP as broadly as possible. To make our protocol definite, we review quantum state preparation procedures for matrix product states (MPS).
Consider a one-dimensional quantum system composed of $N$ qudits, each with local physical dimension $d$. An MPS of bond dimension $\chi$ with open boundary conditions is defined by the amplitude representation~\cite{PerezGarcia2007, Schollwock2011}:
\begin{equation}
    |\Psi\rangle = \sum_{s_1, \dots, s_N = 0}^{d-1} \bra{L} A^{[1] s_1} A^{[2] s_2} \cdots A^{[N] s_N} \ket{R} |s_1 s_2 \dots s_N\rangle,
    \label{eq:mps_def}
\end{equation}
where each $A^{[k] s_k}$ is a $\chi \times \chi$ complex matrix acting on the virtual bond space, and $\bra{L}, \ket{R} \in \mathbb{C}^\chi$ are the virtual boundary vectors. The parameter $\chi$ intrinsically bounds the bipartite entanglement of the state. For mapping to quantum circuits, the MPS is typically cast into left-canonical form, enforcing the isometry condition $\sum_{s} (A^{[k] s})^\dagger A^{[k] s} = \mathbb{I}_\chi$ at each site.

The canonical deterministic algorithm to initialize the state in Eq.~\eqref{eq:mps_def} on a digital quantum processor is \textit{sequential preparation}~\cite{Schon2005}. By exploiting the left-canonical isometry condition, the matrices $A^{[k]}$ can be rigorously embedded into a sequence of unitary operators $U^{[k]}$. Each unitary acts jointly on a physical qudit (initialized in $|0\rangle_p$) and a $\lceil \log_2 \chi \rceil$-qubit ancillary register (the virtual memory, denoted by subscript $v$). The action of $U^{[k]}$ maps an incoming virtual index $\alpha$ to an outgoing index $\beta$ while outputting the required physical state:
\begin{equation}
    U^{[k]} \left( |0\rangle_p \otimes |\alpha\rangle_v \right) = \sum_{s=0}^{d-1} \sum_{\beta=0}^{\chi-1} \left( A^{[k] s} \right)_{\beta \alpha} |s\rangle_p \otimes |\beta\rangle_v.
    \label{eq:sequential_unitary}
\end{equation}
By initializing the ancilla in the boundary state $\ket{L}_v$ and iteratively applying $U^{[1]}, U^{[2]}, \dots, U^{[N]}$ in a staircase circuit architecture, the target MPS is generated exactly. However, this iterative entanglement spreading mandates a quantum circuit depth that scales linearly with the system size, $\mathcal{O}(N)$~\cite{Schon2005, Ran2020}. 

To reduce this $\mathcal{O}(N)$ circuit depth recent works have employed probabilistic mid-circuit measurements and local operations and classical communication (LOCC)~\cite{Piroli2021, Smith2023, Malz2024}. In this paradigm, the global MPS is spatially partitioned into disjoint blocks. These shorter segments are prepared in parallel in $\mathcal{O}(1)$ depth, routing their internal virtual bonds to physical boundary qudits.

Adjacent blocks are subsequently ``stitched'' together via fusion measurements that fuse the right boundary of block $L$ and the left boundary of block $R$ using a Bell-basis measurement (BSM) on the adjacent boundary qudits. Projecting onto a Bell state $|\Phi_m\rangle = (\mathbb{I} \otimes P_m) |\Phi^+\rangle$---where $m$ indexes the classical outcome and $P_m$ is the corresponding Pauli byproduct operator---teleports the virtual entanglement across the partition. This projective measurement successfully contracts the shared virtual bond, and heralds the insertion of a byproduct operator $P_m$ into the tensor network:
\begin{equation}
    \dots A^{[k] s_k}_{L} A^{[k+1] s_{k+1}}_{R} \dots \xrightarrow{\text{BSM } m} \dots A^{[k] s_k}_{L} P_m A^{[k+1] s_{k+1}}_{R} \dots
    \label{eq:fusion_defect}
\end{equation}
The measurement collapses the wavefunction, and outcomes where $P_m \neq \mathbb{I}$ represent heralded Pauli defects in the virtual space.

As in all teleportation protocols these defects must be corrected to recover the target state~\cite{Smith2024}. For specific classes of symmetric or injective MPS (such as symmetry-protected topological phases), a virtual Pauli defect can be algebraically commuted through the local tensors and mapped to a physical unitary correction $V_m$ applied to the physical qudits~\cite{Smith2023, Stephen2024}:
\begin{equation}
    P_m A^{[k] s} = \sum_{s'=0}^{d-1} (V_m)_{s s'} A^{[k] s'} P'_{m}.
    \label{eq:push_through}
\end{equation}
By systematically pushing the byproduct operators to the boundaries of the chain (updating them to $P'_m$ at each site) or actively correcting them via classical feed-forward of the measurement record, the desired state is recovered deterministically. By trading deep coherent evolution for shallow parallel generation and projective entanglement swapping, measurement-based methods compress the required preparation depth from $\mathcal{O}(N)$ to logarithmic $\mathcal{O}(\log N)$~\cite{Malz2024}, and for suitable phases of matter, strictly constant $\mathcal{O}(1)$ depth~\cite{Smith2023, Smith2024, Stephen2024}. For molecular state preparation, and for the Fe-S complexes studied in~\cite{Lee_2023}, sequential state preparation approaches have been optimized for fault tolerant quantum hardware~\cite{berry2025rapid}.

\subsection{Clock Hamiltonian}
\label{sec:clock_hamiltonian}
A Clock Hamiltonian is constructed on a bipartite Hilbert space $\mathcal{H}_C = \mathbb{C}^{T+1} \otimes \mathcal{H}_{\text{sys}}$, where $\mathbb{C}^{T+1}$ corresponds to the space of the counter (clock register) and $\mathcal{H}_{\text{sys}}$ is the $n$-qubit computational space that the circuit acts upon. The Clock Hamiltonian encodes the entire computational history of a quantum circuit into its ground state, known as the \textit{history state}. Let $|\psi_t\rangle$ denote the state of the computational register after the $t^\text{th}$ gate has been applied, and let $|t\rangle$ be the corresponding computational basis state of the ancillary clock register. The history state is defined as:
\begin{equation}
    |\Psi_{\text{hist}}\rangle = \frac{1}{\sqrt{T+1}} \sum_{t=0}^{T} |t\rangle \otimes |\psi_t\rangle,
\end{equation}
where $T$ is the total number of computational steps. 

The Clock Hamiltonian is a sum of initialization and propagation terms:
\begin{equation}
    H_{\text{clock}} = H_{\text{init}} + H_{\text{prop}}.
\end{equation}
The term $H_{\text{init}}$ enforces an energetic penalty for any computational qubit that is not correctly initialized to $|0\rangle$ when the clock is at $t=0$:
\begin{equation}
    H_{\text{init}} = |0\rangle\langle0|_{\text{clock}} \otimes (I_{\text{sys}} - |0\dots0\rangle\langle0\dots0|),
\end{equation}
where $\Pi^{(0)}_i$ is the projector onto the $|0\rangle$ state for the $i$-th qubit. In proofs of QMA- completeness, an additional penalty term $H_{out}= |T\rangle\langle T|\otimes |0\rangle\langle0|_{out}$ is added to ensure that the ground state represents only evolutions on which a particular qubit takes the value one. We do not make this restriction here and so omit this term.

The propagation term $H_{\text{prop}}$ governs the forward and backward dynamics of the computation:
\begin{equation}
    H_{\text{prop}} = \sum_{t=0}^{T-1} H_{t,t+1},
\end{equation}
with each local term defined as:
\begin{equation}
\begin{aligned}
    H_{t,t+1} = \frac{1}{2} \Big( &|t\rangle\langle t| \otimes \mathbb{I} + |t+1\rangle\langle t+1| \otimes \mathbb{I} \\
    &- |t+1\rangle\langle t| \otimes U_{t+1} - |t\rangle\langle t+1| \otimes U_{t+1}^\dagger \Big),
\end{aligned}
\end{equation}
where $U_{t+1}$ is the unitary operator applied at the $(t+1)^\text{th}$ step of the circuit (with $U_0 \equiv \mathbb{I}$). 

It is an interesting property of the clock Hamiltonian that it's spectrum is independent of the details of the circuit whose history it encodes, depending only on the circuit depth. One can define a basis transformation unitary $W$ that isolates the computational evolution:
\begin{equation}
    W = \sum_{t=0}^{T} |t\rangle\langle t| \otimes \mathcal{U}_t,
\end{equation}
where $\mathcal{U}_t = U_t U_{t-1} \dots U_1$. By moving into this clock basis, the Hamiltonian transforms as $\tilde{H} = W^\dagger H_{\text{clock}} W$. The initialization term is largely unaffected ($\tilde{H}_{\text{init}} = H_{\text{init}}$), while the propagation Hamiltonian takes the form:
\begin{equation}
    \tilde{H}_{\text{prop}} = H_{\text{kin}}\otimes \mathbb{I}.
\end{equation}
The matrix $H_{\text{kin}}$ takes the form of a discrete 1D Laplacian:
\begin{equation}
\begin{split}
    H_{\text{kin}} &= \frac{1}{2}\sum_{t=0}^{T-1} \left(|t\rangle\langle t| + |t+1\rangle\langle t+1| - |t\rangle\langle t+1| - |t+1\rangle\langle t|\right) \\
    &= \begin{bmatrix}
    \frac{1}{2} & -\frac{1}{2} & 0 & \dots & 0 \\
    -\frac{1}{2} & 1 & -\frac{1}{2} & \dots & 0 \\
    0 & -\frac{1}{2} & 1 & \dots & 0 \\
    \vdots & \vdots & \vdots & \ddots & \vdots \\
    0 & 0 & 0 & \dots & \frac{1}{2}
    \end{bmatrix}.
\end{split}
\end{equation}
This indicates that the clock propagation is dynamically equivalent to the discretization of a free particle moving on a 1D discrete line ($-\partial^2_x/2$). 

If we were to measure the clock register, the history state would collapse to a single time slice. The probability of collapsing to any given time slice is $1/(T+1)$. The probability of observing the output of the circuit can be amplified arbitrarily close to $1$ by appending identity operations to the end of the active circuit. we consider a scenario where the state preparation requires an {\em active depth} $D_c$, but the total number of allocated clock steps is $T > D_c$. For time steps $t > D_c$, the circuit applies identity gates, meaning the target solution $|\psi_{\text{target}}\rangle = |\psi_{D_c}\rangle$ remains static. The corresponding probability of observing the target solution becomes $P = (T - D_c + 1)/(T+1)$. This padding is widely used to amplify success probabilities in QMA proofs and is known as idling~\cite{Caha_2018}.

\section{Theory}\label{sec:Theory}
In this section we develop the theoretical basis of Universal Parent Hamiltonian Adiabatic Warm Start (UPHAWS) protocol. First we develop the clock Hamiltonian as a universal parent Hamiltonian. Next we review the gap of the clock Hamiltonian, before extending the clock construction to circuits with probabilistic gates. Finally we give the rotating frame presentation of the full adiabatic Hamiltonian that will be the basis of the numerical calculations described in section~\ref{sec:numerical_results}.

\subsection{Clock Hamiltonian as a Parent Hamiltonian}
\label{sec:parent_hamiltonian}

Based on the observation that idling the circuit naturally amplifies the measurement probability of the final state, we split the history state into the active preparation phase and the idling phase:
\begin{equation}
    |\Psi_{\text{hist}}\rangle = \frac{1}{\sqrt{T+1}} \left( \sum_{t=0}^{D_c} |t\rangle \otimes |\psi_t\rangle + \sum_{t=D_c+1}^{T} |t\rangle \otimes |\psi_{\text{target}}\rangle \right).
\end{equation}
We compute the overlap with a reference state $|\tilde{\Psi}_g\rangle$, defined as the tensor product of the uniform superposition state of the clock register with the target state:
\begin{equation}
    |\tilde{\Psi}_g\rangle = \left( \frac{1}{\sqrt{T+1}} \sum_{t=0}^{T} |t\rangle \right) \otimes |\psi_{\text{target}}\rangle.
\end{equation}
The overlap $\langle \tilde{\Psi}_g | \Psi_{\text{hist}} \rangle$ is 
\begin{align}
    \langle \tilde{\Psi}_g | \Psi_{\text{hist}} \rangle 
    &= 1 - \frac{D_c+1}{T+1} \left( 1 - \overline{\langle \psi_{\text{target}} | \psi_t \rangle} \right),
\end{align}
where we have defined the average overlap during the active preparation phase as 
\begin{align}
    \overline{\langle \psi_{\text{target}} | \psi_t \rangle} = \frac{1}{D_c+1} \sum_{t=0}^{D_c} \langle \psi_{\text{target}} | \psi_t \rangle.
\end{align}

Padding the total clock length $T$ while holding the preparation depth $D_c$ fixed drives this overlap to $1$ with cost $O(\frac{D_c}{T})$. Since any quantum state with a polynomial-depth preparation circuit can be embedded by this construction, the clock Hamiltonian serves as a local parent Hamiltonian for any such state.

We use this construction as the starting point of a warm-started adiabatic protocol. The adiabatic Hamiltonian interpolates between the static clock Hamiltonian and a target term that combines an energetic penalty on the clock register with the problem Hamiltonian acting on the system register (Section \ref{sec:theoretical_framework}). 
\begin{equation*}
    \begin{split}
        H_A(s) = &(1-s)H_{\text{clock}} \\
                 &+ s \left((I-|+\rangle\langle+|_c) \otimes I_{\text{sys}} + I_c \otimes H_{\text{target}}\right).
    \end{split}
\end{equation*}
In the following subsection, we analyze the sepctral gap of this initial Hamiltonian. 

\subsection{Spectral Gap of the Clock Hamiltonian}
\label{sec:theoretical_framework}

In UPHAWS, the clock Hamiltonian $H_{\text{clock}}$ sets the initial point of the adiabatic path at $s=0$, and its spectral gap controls the initial gap of the warm-started ASP problem. The minimum gap along the full path depends on the target Hamiltonian and the interpolation, and is studied numerically in Sec.~\ref{sec:numerical_results}. The gap of $H_{\text{clock}}$ is well known, but to fix notation and to keep the paper self contained we compute it here.

Consider a quantum circuit for state preparation with no ancilla of depth $D_c$ which prepare a quantum state deterministically, where $D_c$ counts the sequential gate layers after parallelizable gates are grouped. The clock register has length $T+1$ with $T \ge D_c$; slots $t > D_c$ are padded with identity operations so that the history state retains $|\psi_{\text{target}}\rangle$ in its system register at every padded slot. In the rotated frame defined by
\begin{equation}\label{eqn:W}
    W = \sum_{t=0}^{T} |t\rangle\langle t| \otimes \mathcal{U}_t, \qquad \mathcal{U}_t = U_t U_{t-1} \cdots U_1,
\end{equation}
the propagation term becomes time translation-invariant:
\begin{equation}
    W^\dagger H_{\text{prop}} W = H_{\text{kin}} \otimes I_{\text{sys}},
\end{equation}
where $H_{\text{kin}}$ is the discrete graph Laplacian on the 1D lattice of $T+1$ clock sites. Its eigenvalues are
\begin{equation}
    \epsilon_k = 1 - \cos(\omega_k) \approx \frac{2\pi^2 k^2}{(T+1)^2}, \qquad \omega_k = \frac{2\pi k}{T+1},
\end{equation}
so the spacing between the ground mode ($k=0$) and the first excited mode ($k=1$) is $\Theta(T^{-2})$. The initialization penalty acts on a single clock slot and contributes off-diagonal couplings of magnitude $\mathcal{O}(1/(T+1))$ between momentum sectors. These couplings shift the low-lying spectrum without modifying the $\Theta(T^{-2})$ scaling, so
\begin{equation}
    \Delta(H_{\text{clock}}) = \Theta(T^{-2}).
\end{equation}

The history state has overlap $1 - \mathcal{O}(D_c/T)$ with $|\tilde{\Psi}_g\rangle$, so the minimum clock length needed for an order-unity overlap is $T = \Theta(D_c)$. For a deterministic preparation circuit, the gap of the initial clock Hamiltonian therefore scales as
\begin{equation}
    \Delta(H_{\text{clock}}) = \Theta(D_c^{-2})
\end{equation}
in the active circuit depth. The full rotated Hamiltonian along the path, including the target term, and its momentum-basis matrix elements, are given in Appendix~\ref{app:momentum_basis}.

\subsection{Probabilistic Preparation via Amplitude Amplification}
\label{sec:probabilistic_preparation}

The preparation circuit underlying the clock Hamiltonian is assumed deterministic in the analysis above. Now consider the case when the preparation circuit is probabilistic---succeeding with probability $p_{\text{success}} = \sin^2\theta < 1$ conditional on a heralded flag measurement.

Let $\mathcal{A}$ denote the probabilistic preparation circuit on $n$ flag qubits and $m$ system qubits,
\begin{equation}
    \mathcal{A}\, |0\rangle^{\otimes n} |0\rangle^{\otimes m} = \sin\theta \bigl(|0\rangle^{\otimes n} \otimes |\phi\rangle\bigr) + \cos\theta\, |\Phi^\perp\rangle,
\end{equation}
with $|\phi\rangle$ the target system state. 

$A$ is a unitary operator on $m$ clock-state qubits and $n$ flag qubits, encoding a circuit replacing some $U_t$ by $A$ and expanding the system register. However, this will result in a final state that is a superposition of correct ($|0\rangle^{\otimes n}$) and incorrect flag qubit states. The probability of observing all flag states zero will shrink exponentially with $n$. Idling will not increase the probability of observing correct flag states. Thus at most $\mathcal{O}(\log n)$ probabilistic gates are allowed in an efficient protocol. Applying energy penalties to the flag qubits will make the spectrum of $H_{\text{clock}}$ depend on the details of the circuit, thus making the gap analysis no longer straightforward.

However, we can decrease the probability of observing $|\Phi^\perp\rangle$ within the circuit itself by using amplitude amplification. This can only polynomially increase the probability of success, but serves to ensure that the final state is close to $|\psi_{\text{target}}\rangle \otimes |0\rangle_{\text{flag}} \otimes |\text{clock}\rangle$.

Iterating the Grover operator $Q = -R_\Psi R_{\text{good}}$, with $R_\Psi = I - 2\mathcal{A}|0\rangle\langle 0|\mathcal{A}^\dagger$ and $R_{\text{good}} = I - 2(|0\rangle\langle 0|^{\otimes n} \otimes I^{\otimes m})$, rotates the success amplitude to $\sin\bigl((2K+1)\theta\bigr)$. Setting $K = \lceil \pi/(4\theta) \rceil$ drives this to order unity at depth overhead $\mathcal{O}(1/\sqrt{p_{\text{success}}})$. Fixed-point amplitude amplification~\cite{Yoder_2014} achieves the same bound when $p_{\text{success}}$ is not known in advance.

The amplified circuit $Q^K \mathcal{A}$ has effective active depth
\begin{equation}
    D_{\text{total}} = (2K+1) D_c = \Theta\!\left(D_c / \sqrt{p_{\text{success}}}\right).
\end{equation}
Treating the amplified circuit as deterministic and applying the result of the previous subsection gives the gap of the corresponding clock Hamiltonian:
\begin{equation}
    \Delta(H_{\text{clock}}) = \Theta\!\left(\frac{p_{\text{success}}}{D_c^2}\right).
\end{equation}
For preparation circuits whose native depth $D_c(n)$ and success probability $p_{\text{success}}(n)$ scale polynomially with the system size, the initial gap of the warm-started adiabatic path is polynomially bounded.

To contextualize the scaling constraints of probabilistic state preparation, consider a generic protocol that relies on the conditional measurement of $f$ independent flag qubits. If each flag yields the desired outcome with probability $p < 1$, the global success probability prior to amplification decays exponentially with the number of flags, $p_{\text{success}} = p^f$. For many-body preparation schemes where the number of flags scales extensively with the system size ($f \propto N$), overcoming this exponential suppression via amplitude amplification requires an exponentially deep circuit overhead of $\mathcal{O}(p^{-f/2})$. As established in Eq.~(26), encoding such an amplified circuit into the clock Hamiltonian would cause its spectral gap to vanish exponentially, $\Delta(H_{\text{clock}}) = \mathcal{O}(p^f/D_c^2)$, thereby rendering the adiabatic warm-start computationally intractable.

However, the probabilistic matrix product state preparation circumvent this extensive spatial bottleneck. When employing sequential unitary dilations, the bulk site tensors are applied deterministically. The probabilistic overhead arises exclusively from a finite number of boundary projections---specifically, the generalized Bell-basis measurement required to trace out the virtual bond register. Because the correctly projected boundary state carries an exact amplitude factor of $1/D$, the unamplified global success probability scales polynomially in bond dimension as $p_{\text{success}} = \Theta(1/D^2)$. 

This success probability is decoupled from the physical system size $N$ and depends only on bond dimension $D$. It is the dependence of bond dimension on system size that determines the scaling of the initial gap. To drive this probability to near-unity, amplitude amplification requires only a tractable number of Grover iterations, scaling linearly with the bond dimension as $\mathcal{O}(D)$. Substituting this success probability into our gap analysis yields an initial spectral gap of:
\begin{equation}
    \Delta(H_{\text{clock}}) = \Omega\left(\frac{1}{D^2 D_c^2}\right).
\end{equation}
Because the required amplification overhead introduces only a polynomial factor, the initial spectral gap of the encoded initial clock Hamiltonian is bounded from below. This theoretical guarantee ensures that the Universal Parent Hamiltonian Adiabatic Warm Start (UPHAWS) framework can efficiently embed  MPS with bond dimension polynomial in system size without suffering from the drwabacks of generic post-selected quantum circuits. To go beyond bond dimension polynomial in system size requires sequential state preparation circuits without intermediate measurements whose depth scales linearly with the number of qubits~\cite{schon2005sequential}.

\subsection{Rotating-Frame Construction and Momentum-Space Truncation}\label{sec:rotframe}

The clock register multiplies the Hilbert-space dimension by $T+1$, increasing the cost of direct simulation of the warm-started adiabatic evolution. We instead work in a rotating frame that diagonalizes the clock propagation, write down the effective Hamiltonian in terms of blocks indexed by clock momentum. We then truncate the basis to the slow modes.

The rotation is implemented by the unitary W defined in~\ref{eqn:W}. Conjugating each term in $H_A(s)$  by $W$ gives:
\begin{align}
    W^\dagger H_{\text{prop}} W &= H_{\text{kin}} \otimes I_{\text{sys}}, \\
    W^\dagger H_{\text{in}} W &= |0\rangle\langle 0|_c \otimes \Pi_{\text{init}}, \\
    W^\dagger (|+\rangle\langle+|_c \otimes I_{\text{sys}}) W &= \frac{1}{T+1}\sum_{t,\tau=0}^{T} |t\rangle\langle\tau| \otimes \mathcal{U}_t^\dagger \mathcal{U}_\tau, \\
    W^\dagger (I_c \otimes H_{\text{target}}) W &= \sum_{t=0}^{T} |t\rangle\langle t| \otimes \mathcal{U}_t^\dagger H_{\text{target}} \mathcal{U}_t,
\end{align}
where $H_{\text{kin}}$ is the discrete Laplacian on the 1D clock path, $\Pi_{\text{init}} = I_{\text{sys}} - |0\dots 0\rangle\langle 0\dots 0|$, and we have used $\mathcal{U}_0 = I$. If we write $\mathcal{U}_t^\dagger H_{\text{target}} \mathcal{U}_t$ as a local Hamiltonian in the basis of pauli operators the number of terms could grow exponentially with $D_c$. Instead we build each sub matrix directly which has size as large as $H_{target}$. This accomplishes the goal of achieving a direct classical simulation with cost scaling exponentially only in the number of system qubits.

Projecting the rotated Hamiltonian onto the clock momentum states
\begin{equation}
    |k\rangle = \frac{1}{\sqrt{T+1}}\sum_{t=0}^{T} e^{i \omega_k t} |t\rangle, \qquad \omega_k = \frac{2\pi k}{T+1},
\end{equation}
gives an effective Hamiltonian indexed by momentum pairs $(k,j)$, with each block acting on the system register. Using $\langle k|t\rangle = (T+1)^{-1/2} e^{-i \omega_k t}$ and the rotated forms above, the blocks are
\begin{align}
    M_{kj}(s) ={}& \delta_{kj}\bigl[(1-s)\epsilon_k + s\bigr] I_{\text{sys}} + \frac{1-s}{T+1} \Pi_{\text{init}} \nonumber \\
    &- \frac{s}{(T+1)^2}\, \mathcal{V}_k \mathcal{V}_j^\dagger \nonumber \\
    &+ \frac{s}{T+1}\sum_{t=0}^{T} e^{-i(\omega_k - \omega_j) t}\, \mathcal{U}_t^\dagger H_{\text{target}} \mathcal{U}_t,
\end{align}
where $\epsilon_k = 1 - \cos(\omega_k) \approx 2\pi^2 k^2/(T+1)^2$ and $\mathcal{V}_k \equiv \sum_{t=0}^{T} e^{-i \omega_k t}\, \mathcal{U}_t^\dagger$ is the discrete Fourier transform of the cumulative circuit. The diagonal blocks ($k=j$) carry the clock kinetic energy and a uniform shift from the target penalty; the off-diagonal blocks couple momentum sectors through the initialization penalty, the rotated target Hamiltonian, and $\mathcal{V}_k$. For the cv detailed derivation see Appendix~\ref{app:momentum_basis}.

The diagonal energy $\epsilon_k$ grows quadratically in $k$, so the slow modes ($k \ll T$) sit at energies $\mathcal{O}(k^2/T^2)$ while the fast modes ($k$ comparable to $T$) sit at energies of order unity. The low-energy adiabatic states, whose energies are at most a few times $\Delta(H_{\text{clock}}) = \Theta(T^{-2})$, are therefore supported on the slow-mode sector of the clock. The off-diagonal blocks coupling slow modes to fast modes carry prefactors $\mathcal{O}(1/T)$ from the initialization penalty and $\mathcal{O}(D_c/T)$ from the rotated target operators, we conjuncture that both will be small compared to the fast-mode energies $\mathcal{O}(1)$. Restricting the simulation to the lowest $k_{\text{cut}}$ momentum modes therefore captures the physics that determines the gap, with the discarded subspace acting on the retained one only through these weak, off-diagonal couplings.

    \section{Numerical Results}
\label{sec:numerical_results}

We use numerical simulation to test three aspects of the UPHAWS. First, we study the adiabatic spectral gap along the interpolation from a clock Hamiltonian encoding a $\mathbb{Z}_2$-symmetric MPS preparation circuit to a target GHZ Hamiltonian, varying the warm-start parameter and the system size. Second, we benchmark the momentum-space truncation scheme that makes simulation of the augmented clock space tractable. Third, we apply the framework to a strongly correlated quantum chemistry benchmark, the linear $H_6$ chain under symmetric bond stretching, and compare against adiabatic state preparation from the Hartree--Fock state.

\subsection{$\mathbb{Z}_2$-Symmetric MPS to GHZ}

Generating the highly entangled Greenberger-Horne-Zeilinger (GHZ) state via ASP typically forces the system across a topological phase boundary, precipitating an exponentially vanishing spectral gap, making ASP inefficient~\cite{sachdev1999quantum,Koma_1994}. To circumvent this finite-size scaling bottleneck, the preparation pathway must be carefully restricted to a symmetry-protected subspace. We achieve this by warm-starting the evolution using a $\mathbb{Z}_2$-symmetric family of translationally invariant Matrix Product States (MPS) with bond dimension $D=2$, previously introduced by Wolf \textit{et al.}~\cite{wolf2006quantum}. Each state in this family is parameterized by a real parameter $g \in [-1,0]$, with onsite tensors generated by:
\begin{equation}
    A^0 = \eta
    \begin{pmatrix}
    1 & 0\\[6pt]
    \sqrt{-g} & 0
    \end{pmatrix},
    \quad
    A^1 = \eta
    \begin{pmatrix}
    0 & -\sqrt{-g}\\[6pt]
    0 & 1
    \end{pmatrix}
\end{equation}
where $\eta = \frac{1}{\sqrt{1 - g}}$.At $g=-1$, the MPS reduces to a trivial product state. However, as $g \to 0$, it continuously approaches the highly entangled $n$-qubit GHZ state.

The generalized $n$-qubit GHZ target state is defined as:
\begin{equation}
    |\text{GHZ}_n\rangle = \frac{1}{\sqrt{2}} \left( |0^{\otimes n}\rangle + |1^{\otimes n}\rangle \right).
\end{equation}
We use the standard stabilizer Hamiltonian as the target Hamiltonian for our ASP:
\begin{equation}
    H_{\mathrm{GHZ}_n} = -\Bigl(\sum_{i=0}^{n-2} Z_i Z_{i+1} \;+\; X_0 X_1 \cdots X_{n-1}\Bigr).
\end{equation}
The nearest-neighbor $ZZ$ terms enforce classical ferromagnetic alignment, while the global $X^{\otimes n}$ operator stabilizes the coherent superposition. The intrinsic gap above the ground state of this Hamiltonian is $2$, independent of the system size $n$.

Following the construction of Section \ref{sec:theoretical_framework}, the adiabatic path is
\begin{align}
    H_A(s) &= (1 - s)\,H_{\text{clock}} \nonumber\\
    &\quad + s\biggl[ (I_{\text{clock}} - |+\rangle\langle+|_c) \otimes I_{\text{sys}}\\
    &\phantom{=} + I_{\text{clock}} \otimes H_{\text{GHZ}_n} \biggr],
\end{align}
for $s \in [0,1]$, where $H_{\text{clock}}$ encodes the circuit preparing the $\mathbb{Z}_2$-symmetric MPS from $|0^{\otimes n}\rangle$. We diagonalize $H_A(s)$ on up to five physical data qubits (with the clock and flag ancillae) and extract the minimum spectral gap $\Delta_{\min} = \min_s [E_1(s) - E_0(s)]$ over the path.

\begin{figure}[htbp]
    \centering
    \includegraphics[width=\linewidth]{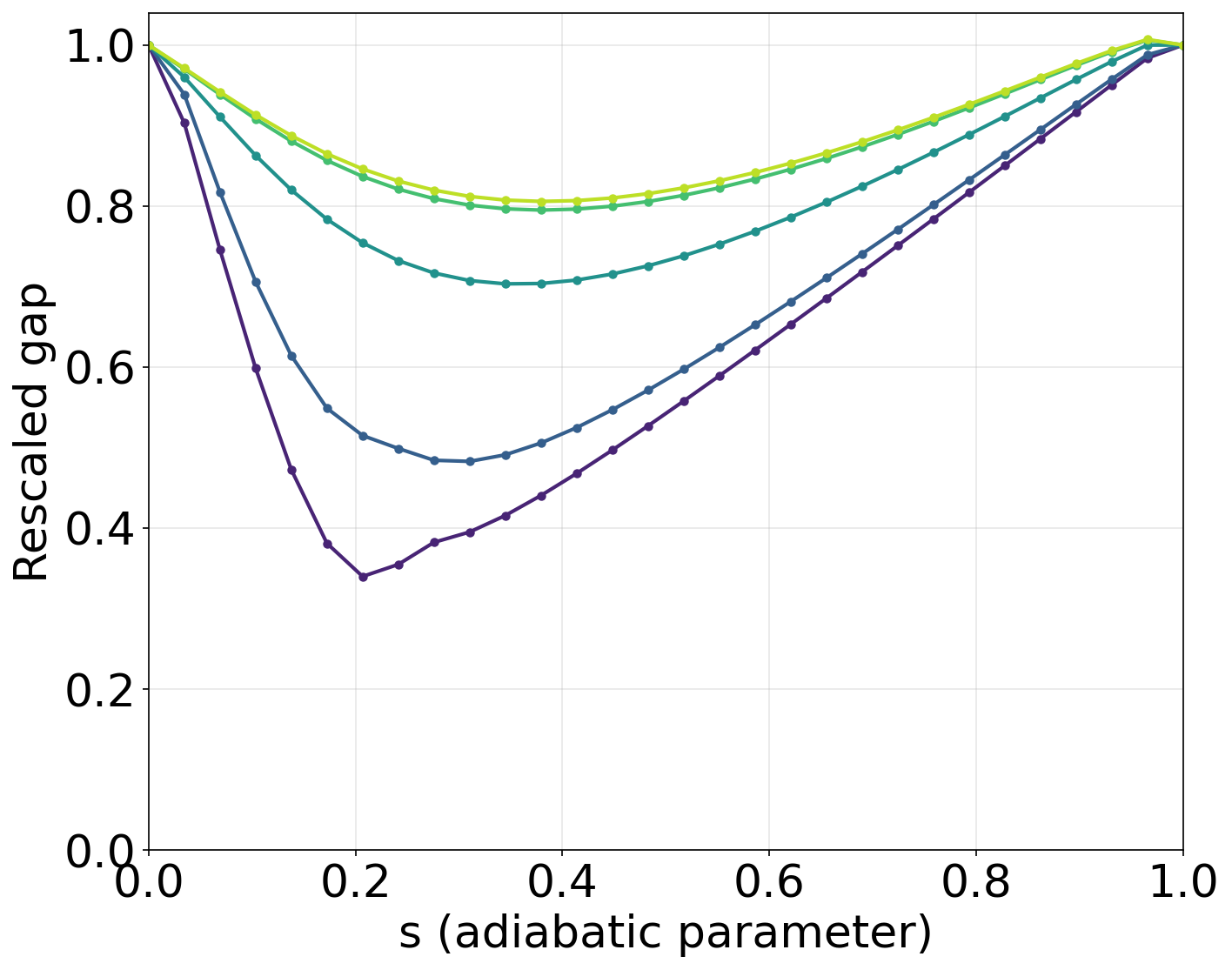}
    \caption{Rescaled spectral gap along the adiabatic path, for several values of the warm-start parameter $g$. From lowest to highest, curves correspond to warm-start values $g=-0.9$ (bottom, dark purple), $-0.5$ (blue), $-0.1$ (teal), $-0.01$ (green), and $-0.001$ (top, yellow), with $\eta = 1/\sqrt{1-g}$.As $g\to0^{-}$ the minimum gap grows from $\Delta_{\min}\approx0.34$ (at $s\approx0.21$) to $\approx0.81$ (at $s\approx0.38$).}
    \label{fig:ghz_gap}
\end{figure}

Figure~\ref{fig:ghz_gap} shows the gap $E_{1}(s)-E_{0}(s)$ between the two lowest eigenvalues of $H(s)$, whose minimum over $s$ controls the adiabatic runtime $\tau\sim\Delta_{\min}^{-2}$. The endpoints of the interpolation $H(s)$ are shifted and scaled so the rescaled gap is unity at $s=0,1$. For $g=-0.9$ and $g=-0.5$ the initial state of the parent Hamiltonian is a trivial product state which, in the limit of large system size, we would expect to exhibit an orthogonality catastrophe. The adiabatic gap for these couplings exhibits a pronounced minimum consistent with inefficient ASP for large systems. Initializing with parameter values closer to the target state results in smoother behavior of the adiabatic gap.

\begin{figure}[htbp]
    \centering
    \includegraphics[width=\linewidth]{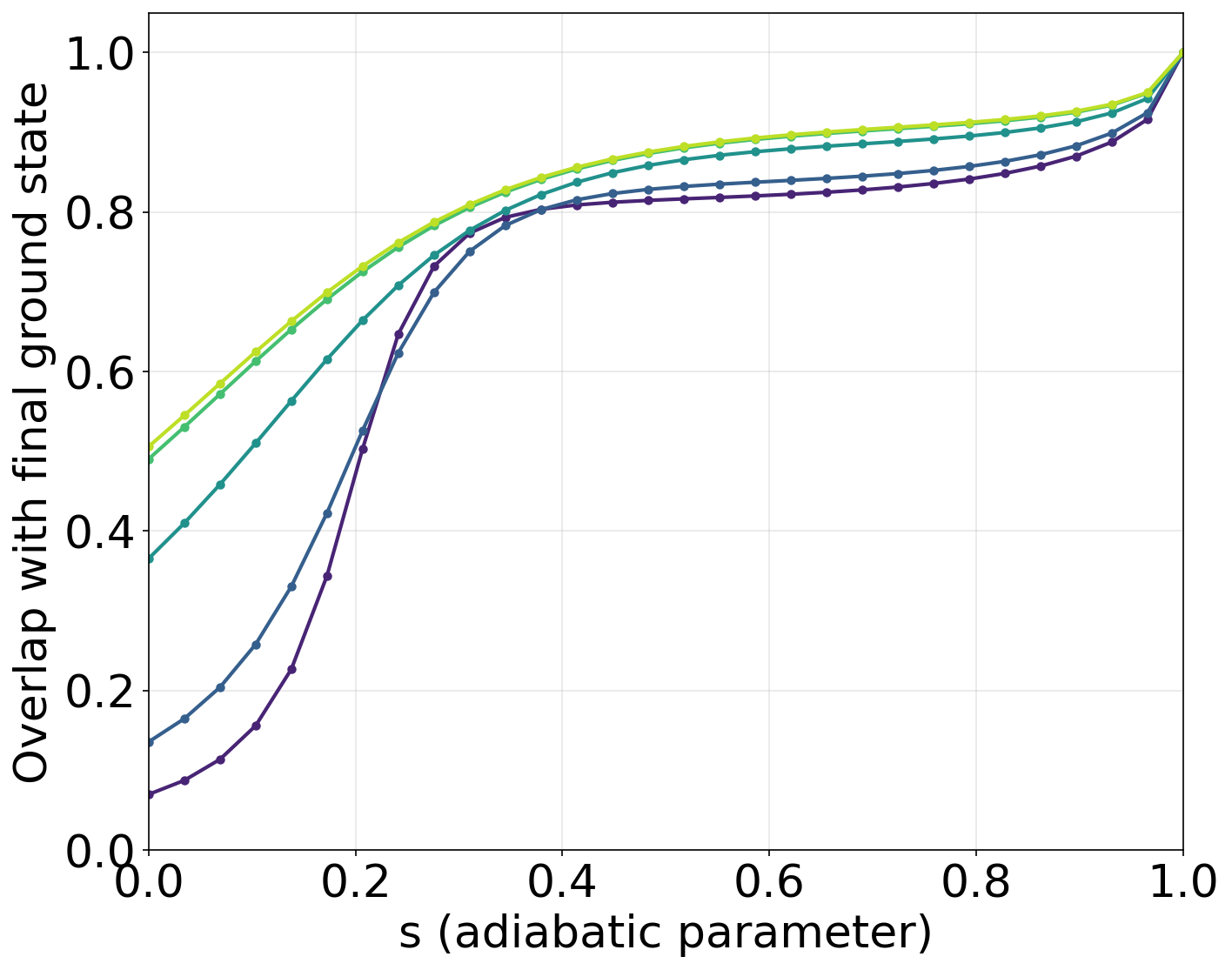}
    \caption{Ground-state overlap with the target GHZ state along the adiabatic path, for several values of the warm-start parameter $g$. Along the interpolation $H(s) = (1-s)\,H_{\mathrm{clock}}/\Delta_{\mathrm{clock}} + s\,H_{\mathrm{final}}/\Delta_{\mathrm{final}}$ ($s\in[0,1]$, each endpoint normalized by its spectral gap $\Delta$), we plot the squared fidelity $|\langle\psi_{0}(s)|\psi_{\mathrm{g}}\rangle|^{2}$ between the instantaneous ground state and the tensor product of uniform superposition and target GHZ state $|\psi_{\mathrm{GHZ}}\rangle = |\psi_{0}(1)\rangle$. Curves correspond to warm-start values $g=-0.9$ (dark purple), $-0.5$ (blue), $-0.1$ (teal), $-0.01$ (green), and $-0.001$ (yellow), with $\eta = 1/\sqrt{1-g}$. The fidelity reaches unity at $s=1$ for all $g$; the initial overlap at $s=0$ falls from $\approx0.51$ ($g=-0.001$) to $\approx0.07$ ($g=-0.9$), the curves fanning out at small $s$ in step with the smaller minimum gap of the larger-$|g|$ warm starts.}
    \label{fig:ghz_overlap}
\end{figure}

Figure \ref{fig:ghz_overlap} traces the instantaneous ground-state overlap with the target GHZ state along the path. Initial states with $g \to 0$ maintain a higher overlap throughout the evolution and exhibit smoother evolution. For initial states with $g=-0.9$ and $g=-0.5$ the overlap changes rapidly in the vicinity of the minimum gap but more slowly once the region of minimum gap has been traversed. Rapid changes in the ground state overlap as a function of adiabatic parameter are associated with a(generalized) orthogonality catastrophe, consistent with the local nature of the initial state.

\begin{figure}[htbp]
    \centering
    \includegraphics[width=\linewidth]{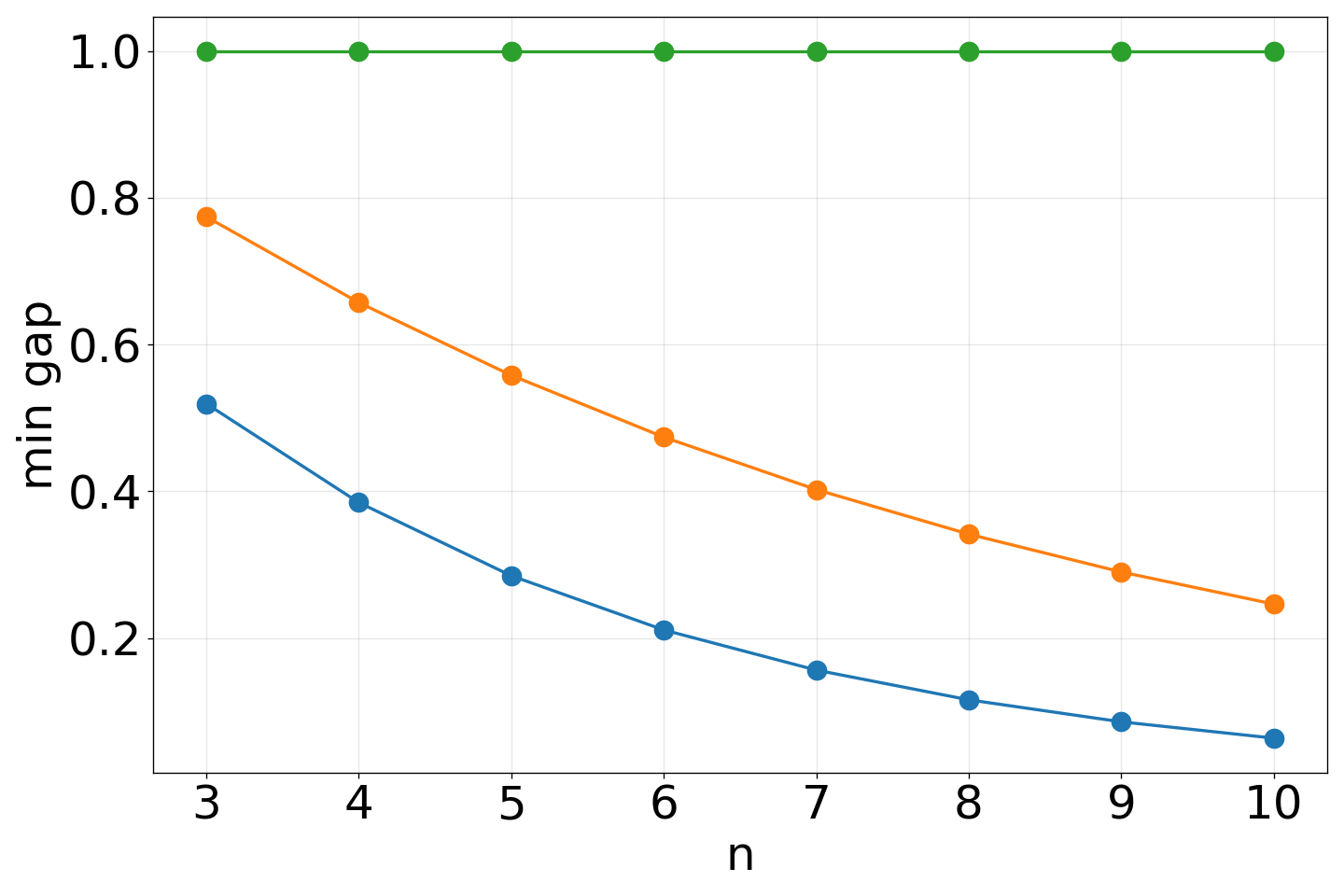}
    \caption{Minimum spectral gap of the warm-start adiabatic interpolation versus the number of system qubits $n$, for three values of the $\mathbb{Z}_2$ matrix-product-state warm-start coupling $g$. We plot $\Delta_{\min} = \min_{s}[E_1(s)-E_0(s)]$. Curves correspond to $g=-1$ (bottom, blue), $g=-0.5$ (middle, orange), and $g=0$ (top, green), with $\eta = 1/\sqrt{1-g}$. The minimum gap decreases with $n$ for $g=-1$ and $g=-0.5$ (to $\Delta_{\min}\approx0.06$ and $\approx0.25$ at $n=10$), whereas $g\to0$ (warm start approaching the GHZ target) keeps $\Delta_{\min}=1$.}
    \label{fig:min-gap-sizes}
\end{figure}

Figure~\ref{fig:min-gap-sizes} shows $\Delta_{\min}$ as a function of $n$ for several values of $g$: initial states closer to the target phase ($g \to 0$) yield larger minimum gaps than the trivial product state at $g = -1.0$.

\subsection{Rotating-Frame Construction and Momentum-Space Truncation}
\label{sec:truncation_numerical}

\begin{figure}[htbp]
    \centering
    \includegraphics[width=\linewidth]{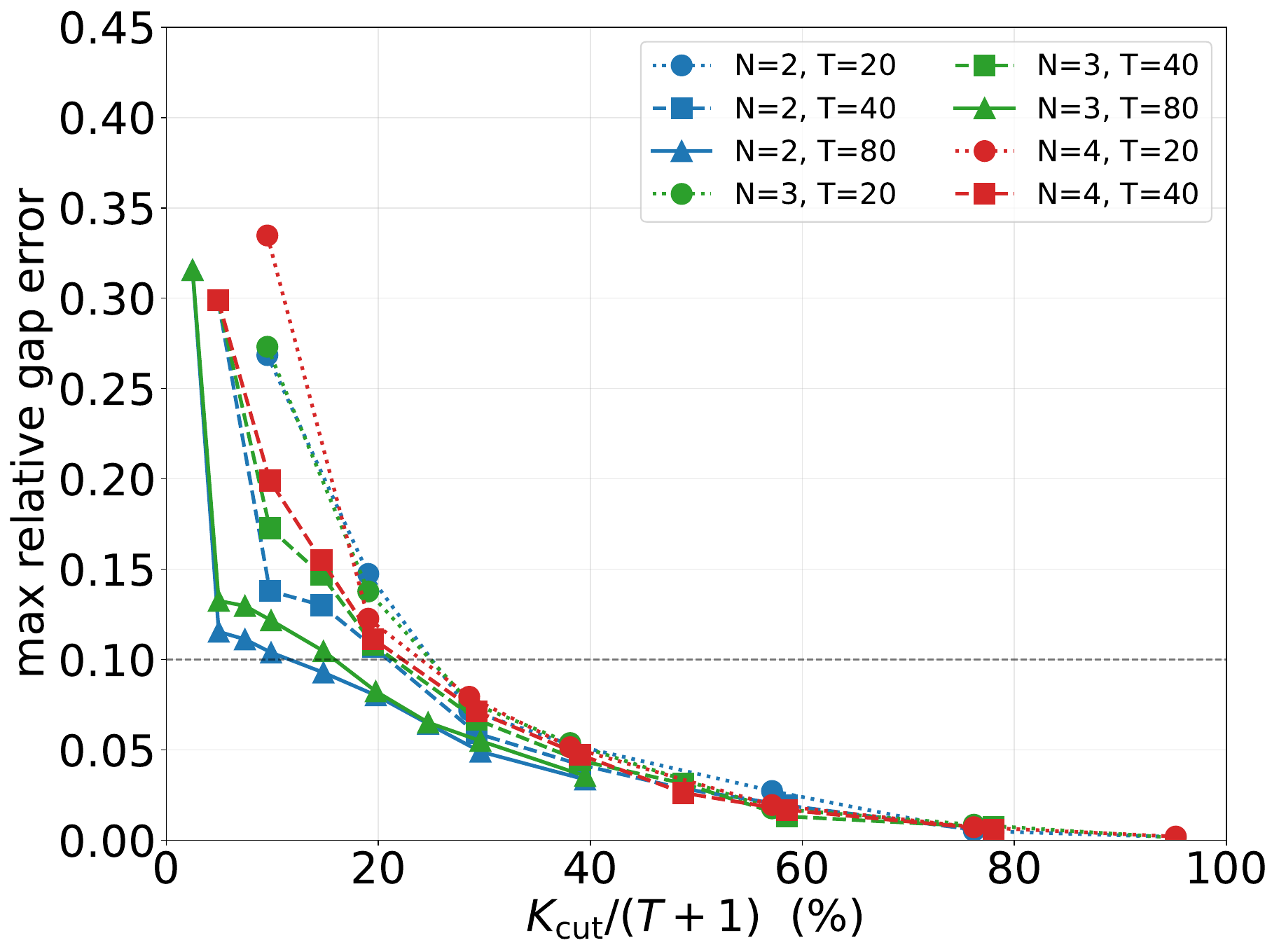}
    \caption{Maximum relative error of the adiabatic spectral gap under momentum-space truncation, as a function of the retained-mode fraction $k_{\text{cut}}/(T+1)$. Curves are shown for system size $n \in \{2,3,4\}$ and clock length $T \in \{20,40,80\}$. The dashed line marks the $10\%$ relative-error level.}
    \label{fig:truncation_benchmark}
\end{figure}

We use our GHZ state preparation to benchmark the rotating frame construction and momentum space truncation introduced in~\ref{sec:rotframe}. We compute the true gap and the gap in the rotating frame construction as a function of truncation level. Figure \ref{fig:truncation_benchmark} reports the maximum relative gap error $\max_s |\Delta_k(s) - \Delta_{\text{ref}}(s)|/\Delta_{\text{ref}}(s)$ as a function of the retained-mode fraction $k_{\text{cut}}/(T+1)$, for system sizes $n \in \{2,3,4\}$ and clock lengths $T \in \{20,40,80\}$. The reference gap $\Delta_{\text{ref}}$ is computed with the full $k_{\text{cut}} = T+1$ basis. The curves show that retaining $20$--$30\%$ of the modes brings the relative error below $10\%$ across all configurations tested, and the error continues to decay as $k_{\text{cut}}$ grows. The accuracy is controlled by the fraction of retained modes rather than their absolute number.

\subsection{Application to Quantum Chemistry: $H_6$ Molecule}
\label{sec:molecular_application}

We apply the UPHAWS to a strongly correlated benchmark from quantum chemistry: ground-state preparation of a linear $H_6$ chain under symmetric bond stretching. The fermionic Hamiltonian is mapped to qubits via the Jordan--Wigner transformation in the STO-3G basis.

Near the equilibrium geometry ($R \lesssim 1.0$~\AA), the ground state is well approximated by a single Slater determinant and the Hartree--Fock (HF) state has substantial overlap with it. As the bonds stretch, the wavefunction acquires multi-reference character and the HF overlap decreases where phase transition happened. Adiabatic state preparation initialized from $\ket{\text{HF}}$ must then traverse a region of small spectral gap, while a warm-start with non-trivial entanglement structure should remain in the same phase as the target.

We compare three protocols on the same target Hamiltonian $H_{\text{mol}}(R)$:
\begin{itemize}
    \item \textbf{HF (ASP):} adiabatic evolution $(1-s) H_{\text{HF}} + s\, H_{\text{mol}}(R)$, with no clock register. The initial Hamiltonian $H_{\text{HF}}$ has the closed-shell HF determinant as its unique ground state.
    \item \textbf{MPS-D2 (clock-warm-start):} the clock Hamiltonian encodes the sequential preparation circuit of a bond-dimension-$D=2$ MPS approximation of the ground state at the same $R$, optimized classically.
    \item \textbf{MPS-D4 (clock-warm-start):} the same construction at $D=4$.
\end{itemize}
For the clock-based protocols, the clock register is padded so the history state has an initial overlap of $0.86$ with $\ket{+}_{\text{clock}} \otimes \ket{\text{MPS}}$. The minimum gap along each path is reported in units of the gap at $s=0$, so that the comparison is not biased by the different Hilbert-space dimensions of the three protocols.

\begin{figure}[htbp]
    \centering
    \includegraphics[width=\linewidth]{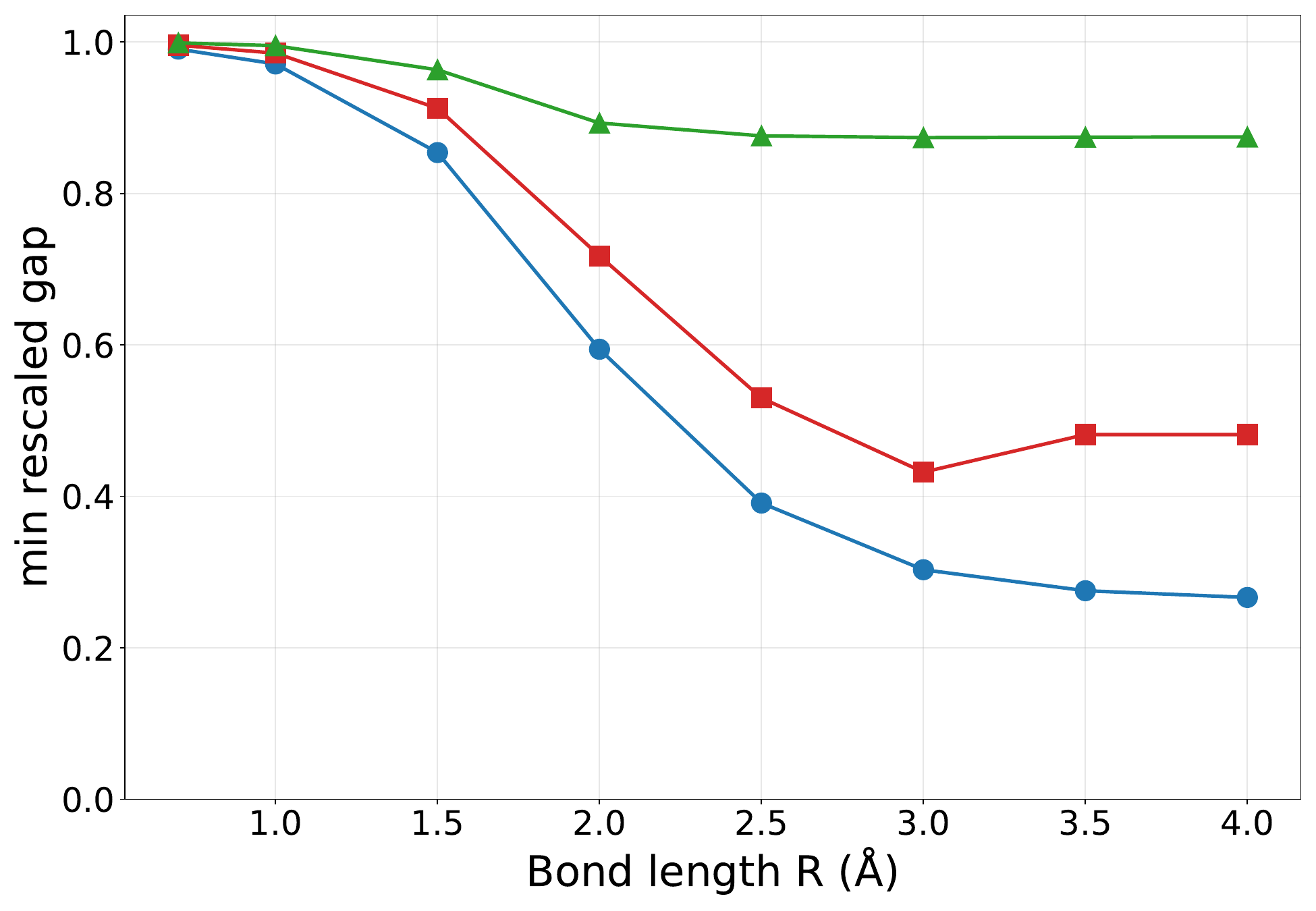}
    \caption{Minimum spectral gap along the adiabatic path for a linear $H_6$ chain under symmetric bond stretching (STO-3G basis, Jordan--Wigner mapping), in units of the gap at $s=0$. HF (blue circles): ASP from the Hartree--Fock state, no clock encoding. MPS-D2 (red squares) and MPS-D4 (green triangles): clock-warm-started from sequential MPS preparation circuits of bond dimension $D=2$ and $D=4$. The clock register is padded so the initial history state has overlap $0.86$ with $\ket{+}_{\text{clock}} \otimes \ket{\text{MPS}}$.}
    \label{fig:molecule-gap}
\end{figure}

Figure \ref{fig:molecule-gap} shows the result. At small atomic separations all three protocols give a comparable gap. As $R$ increases, the HF curve decreases monotonically to $\Delta_{\min}/\Delta(0) \approx 0.27$ at $R = 4.0$~\AA. The minimum gap of the $D=2$ clock-warm-start plateaus near $0.48$ over the same range; wheras that of the $D=4$ clock-warm-start stays above $0.87$. The UPHAWS framework converts warm-start quality, here controlled by the classical bond dimension $D$, into a corresponding adiabatic gap.

\begin{figure}
    \centering
    \includegraphics[width=\linewidth]{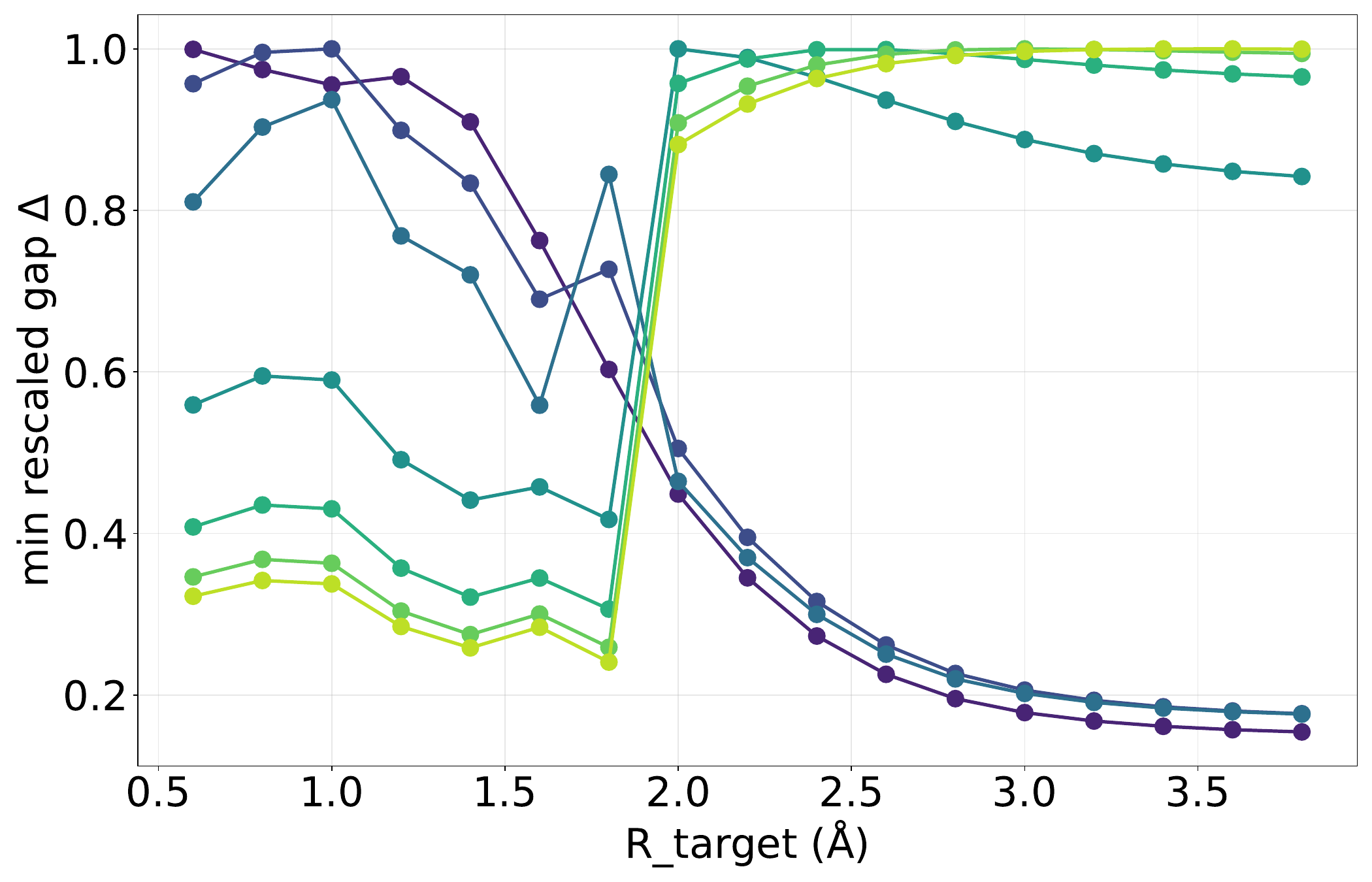}
    \caption{H$_6$ STO-3G---minimum rescaled adiabatic gap each endpoint normalized to unit gap. Each curve uses a different warm-start geometry, in viridis color
 order: $R_{\mathrm{init}}=0.7$ (dark purple), $1.0$ (indigo), $1.5$ (blue),$2.0$ (teal), $2.5$ (green), $3.0$ (yellow-green), and $3.5$~\AA{} (yellow).
  The gap stays open when $R_{\mathrm{init}}$ and $R_{\mathrm{target}}$ lie in
  the same electronic phase and collapses across the phase boundary near
  $R\approx1.9$~\AA{}.}
    \label{fig:H6}
\end{figure}

The results shown in Figure~\ref{fig:molecule-gap} reflect ASP protocols that are designed to succeed. We can also study the failure modes of these protocols, by deliberately choosing the initial state in the opposite phase to that of the target state. We do this by allowing the atomic separation parameter in the warm start state to differ from the atomic separation in the target Hamiltonian. We consider three delocalized/metallic warm starts with $R_{\mathrm{init}}\in\{0.7, 1.0, 1.5\}$~\AA{}, and four localized/tight-binding warm starts with  with $R_{\mathrm{init}}\in\{2.0, 2.5, 3.0, 3.5\}$~\AA{} 
In figure \ref{fig:H6}, we show the minimum gap along the adiabatic path preparing the ground state of H$_6$ as a function of the atomic separation in the target Hamiltonian. The three curves with $R_{\mathrm{init}}\in\{0.7, 1.0, 1.5\}$~\AA{} (delocalized /metallic warm-starts) sit near unity for $R_{\mathrm{target}}\lesssim 1.4$~\AA{} and decay monotonically as $R_{\mathrm{target}}$ stretches into the localized regime, falling below $0.2$ by $R_{\mathrm{target}}\approx 3$~\AA{}. The four curves with $R_{\mathrm{init}}\in\{2.0, 2.5, 3.0, 3.5\}$~\AA{} (localized warm-starts) show the opposite behavior: the gap is below $0.5$ for $R_{\mathrm{target}}\lesssim 1.8$~\AA{}, jumps sharply to near unity at $R_{\mathrm{target}}\approx 2.0$~\AA{}, and stays at $0.85$--$1.00$ throughout the localized region. This clearly indicates that the warm starts succeed in the correct phase, and fail in the incorrect phase.
  
    \section{Conclusions}
\label{sec:conclusion}

In this paper we have defined the UPHAWS protocol for warm-starting ASP using any state that can be efficiently prepared by a quantum circuit. UPHAWS enables ASP to be used in conjunction with any other state preparation scheme. In particular it enables ASP to be warm started with any classical ansatz state. UPHAWS is motivated by a heuristic picture of ASP in which ASP succeeds when it connects two states in the same quantum phase, and fails when it crosses a phase boundary. UPHAWS will be useful in the cases where the quantum phase in which a target state lies is known, where some state in that phase can be described by a classical ansatz or by a state preparation procedure using a short quantum circuit, but where the particular state of interest remains hard to classically describe. In cases where there are $O(1)$ candidate phases UPHAWS can be attempted with each in the same spirit as the proposal of~\cite{PhysRevA.92.062318}. 

The target of quantum state preparation is typically a state with sufficient overlap for ground state estimation on a quantum computer. The target of classical ansatz state development, across the many classical approximate methods, is to develop an ansatz that enables classical computation of the energy. To achieve efficient state preparation quantum algorithms must exploit the same problem structure available to classical methods. An open question in quantum state preparation for chemistry is whether efficient quantum state preparation always implies the existence of a tractable classical ansatz across all of chemical space~\cite{Lee_2023}. UPHAWS allows the warm starting of ASP with any classical ansatz, and hence enables quantum state preparation to improve upon any classical approach. Furthermore UPHAWS enables ASP to be warm-started with any state that admits an efficient preparation by a quantum circuit. Whereas classical ansaetz typically enable classical computation of the energy, a quantum circuit ansatz certainly does not allow efficient classical computation of the energy in general unless $P=BQP$~\cite{lloyd2018quantum,biamonte2021universal,morales2020universality}.

We have demonstrated the construction on two benchmarks: a $\mathbb{Z}_2$-symmetric family of MPS interpolating toward a target GHZ Hamiltonian and the linear $H_6$ chain under symmetric bond stretching. We used the $\mathbb{Z}_2$-symmetric family of MPS to validate the gap scaling and show that initial states closer to the target phase yield larger minimum gaps. For the linear $H_6$ chain under symmetric bond stretching, a bond-dimension-$4$ MPS warm-start encoded as a clock Hamiltonian stabilizes the rescaled minimum gap above $0.87\,\Delta(0)$ across the strongly correlated stretched-bond regime, while ASP from the Hartree--Fock state collapses to about $0.27\,\Delta(0)$ in the same regime. These results indicate the expected success and failure modes of UPHAWS.

In~\cite{Lee_2023}, ASP was simulated from  Hartree Fock and configuration state function initial states. This failure is consistent with our heuristic picture of ASP. Classical MPS approaches were shown to work well for examples where HF initialized ASP failed. Our pictures of phase boundary as obstacles to ASP suggests that starting ASP with a simple MPS in the correct phase should succeed where HF initializations fail. This is a ``warm start'' of ASP: we aim to improve upon existing classical heuristics by directly using them within quantum algorithms. Future work should continue the competition between classical and quantum state preparation methods, including UPHAWS, to achieve improvements in both areas.

A number of interesting questions are left to future work. The analysis of the UPHAWS protocol is complicated by the need for extra qubits to store the clock register. This extra overhead also pushes realization of UPHAWS on fault-tolerant hardware further into the future. A natural direction for future work is the construction of parent Hamiltonians for arbitrary preparation circuits that do not require any additional qubits beyond those of the target system. UPHAWS extends the reach of quantum algorithms for state preparation by enabling ASP to improve upon other classical and quantum ansaetz. The extent to which the patchwork of classical and quantum methods cover the space of interesting systems across quantum many-body physics remains an interesting open question. It is of course likely that new methods will extend the range of systems accessible to simulation, as well as improving the coverage of known systems of interest.

    \section{Acknowledgements}

This work is supported by the NSF STAQ project (PHY-1818914/232580) and by the NSF NQVL:QSTD:Pilot: Quantum Advantage-Class Trapped Ion system (QACTI) project NSF award number 2410675.

\appendix

\section{Derivation of the Effective Hamiltonian in the Momentum Basis}
\label{app:momentum_basis}

In this section, we provide the rigorous, step-by-step derivation of the time-dependent adiabatic Hamiltonian $H_A(s)$ in the momentum basis of the clock register. We employ the interaction picture defined by the isometry $W$ to diagonalize the clock propagation dynamics, allowing us to explicitly isolate the energetic penalty terms responsible for the spectral gap.

\subsection{Interaction Picture Transformation}
To accurately capture the dynamics established in Section \ref{sec:theoretical_framework}, the composite adiabatic Hamiltonian acting on the space $\mathcal{H}_{\text{clock}} \otimes \mathcal{H}_{\text{sys}}$ is given by the decoupled formulation:
\begin{equation}
    H_A(s) = (1-s) H_{\text{clock}} + s \left( |+\rangle\langle+|_c \otimes I_{\text{sys}} + I_c \otimes H_{\text{target}} \right),
\end{equation}
where $|+\rangle = (T+1)^{-1/2} \sum_{t=0}^{T} |t\rangle$ is the uniform superposition of the computational clock states. The static clock Hamiltonian decomposes into initialization and propagation terms, $H_{\text{clock}} = H_{\text{in}} + H_{\text{prop}}$. The initialization term enforces the computational vacuum state $|0\dots0\rangle$ exclusively at time $t=0$:
\begin{equation}
    H_{\text{in}} = |0\rangle\langle 0|_c \otimes \left( \mathbb{I} - |0\dots0\rangle\langle 0\dots0| \right)_s.
\end{equation}
We transition into the interaction frame via the unitary isometry $W = \sum_{t=0}^{T} |t\rangle\langle t| \otimes \mathcal{U}_t$, where $\mathcal{U}_t = U_t \dots U_1$ represents the accumulated computational history up to time step $t$ (with $\mathcal{U}_0 = I$). The effective transformed Hamiltonian is $\tilde{H}(s) = W^\dagger H_A(s) W$.

\subsection{Momentum Basis Representation}
We define the momentum basis states $|k\rangle$ via the Discrete Fourier Transform (DFT) of the temporal clock states:
\begin{equation}
    |k\rangle = \frac{1}{\sqrt{T+1}} \sum_{t=0}^{T} e^{i \omega_k t} |t\rangle, \quad \text{where} \quad \omega_k = \frac{2\pi k}{T+1}.
\end{equation}
We now analytically evaluate the block matrix elements $\mathcal{H}_{kj}(s) \equiv \langle k | \tilde{H}(s) | j \rangle_c$ for each individual component of the Hamiltonian.

\subsubsection{1. Propagation Term}
The term $H_{\text{prop}}$ encodes the discrete kinetic energy of the clock. In the interaction picture, the computational history dependence $\mathcal{U}_t$ is completely untangled and removed by $W$, rendering the propagation operator strictly diagonal in the momentum basis:
\begin{equation}
    \langle k | W^\dagger H_{\text{prop}} W | j \rangle_c = \delta_{kj} \epsilon_k \mathbb{I}_{\text{sys}},
\end{equation}
where $\epsilon_k = 1 - \cos(\omega_k)$ are the scalar eigenvalues of the periodic 1D discrete Laplacian approximation. Note that the lowest energy mode is exactly zero, $\epsilon_0 = 0$.

\subsubsection{2. Initialization Penalty}
The initialization term $H_{\text{in}}$ acts strictly as a local boundary condition at $t=0$. Because $\mathcal{U}_0 = \mathbb{I}$, the transformed operator becomes:
\begin{equation}
    \tilde{H}_{\text{in}} = W^\dagger H_{\text{in}} W = |0\rangle\langle 0|_c \otimes \mathcal{U}_0^\dagger \Pi_{\text{init}} \mathcal{U}_0,
\end{equation}
where $\Pi_{\text{init}} = \mathbb{I} - |0\dots0\rangle\langle 0\dots0|$. Evaluating the matrix element using the overlap identity $\langle k | 0 \rangle_c = (T+1)^{-1/2}$, we obtain a dense, momentum-independent rank-1 contribution across the momentum blocks:
\begin{equation}
    \langle k | \tilde{H}_{\text{in}} | j \rangle_c = \frac{1}{T+1} \tilde{\Pi}_{\text{init}},
\end{equation}
where we have defined $\tilde{\Pi}_{\text{init}} \equiv \mathcal{U}_0^\dagger \Pi_{\text{init}} \mathcal{U}_0 = \Pi_{\text{init}}$.

\subsubsection{3. Target Problem Hamiltonian}
The final adiabatic term penalizes deviations from the target state via both the clock projector and the continuous problem Hamiltonian. The transformed penalty is:
\begin{equation}
    \tilde{H}_{\text{target}} = s W^\dagger \left( |+\rangle\langle+|_c \otimes I_{\text{sys}} + I_c \otimes H_{\text{target}} \right) W.
\end{equation}
By expanding $W$ and using the identity $\langle t | + \rangle = (T+1)^{-1/2}$, we obtain a double sum over the history states for the clock projector, and a single sum for the target Hamiltonian:
\begin{equation}
    \tilde{H}_{\text{target}} = \frac{s}{T+1} \sum_{t, \tau=0}^{T} |t\rangle\langle \tau| \otimes \mathcal{U}_t^\dagger \mathcal{U}_\tau + s \sum_{t=0}^T |t\rangle\langle t| \otimes \mathcal{U}_t^\dagger H_{\text{target}} \mathcal{U}_t.
\end{equation}
Projecting this operator onto the momentum basis evaluates the Fourier transform on both terms:
\begin{equation}
\begin{split}
    \langle k | \tilde{H}_{\text{target}} | j \rangle_c &= \frac{s}{(T+1)^2} \sum_{t, \tau=0}^{T} e^{-i \omega_k t + i \omega_j \tau} \mathcal{U}_t^\dagger \mathcal{U}_\tau \\
    &\quad + \frac{s}{T+1}\sum_{t=0}^T e^{-i(\omega_k - \omega_j)t} \mathcal{U}_t^\dagger H_{\text{target}} \mathcal{U}_t.
\end{split}
\end{equation}
By defining the Fourier-transformed circuit evolution operator as $\mathcal{V}_k \equiv \sum_{t=0}^{T} e^{-i \omega_k t} \mathcal{U}_t^\dagger$, the double sum cleanly factorizes. This simplifies to the effective coupling matrix element:
\begin{equation}
\begin{split}
    \langle k | \tilde{H}_{\text{target}} | j \rangle_c &= \frac{s}{(T+1)^2} \mathcal{V}_k \mathcal{V}_j^\dagger \\
    &\quad + \frac{s}{T+1}\sum_{t=0}^T e^{-i(\omega_k - \omega_j)t} \mathcal{U}_t^\dagger H_{\text{target}} \mathcal{U}_t.
\end{split}
\end{equation}

Combining the propagation, initialization, and final Hamiltonian terms, the full effective block matrix element for the adiabatic Hamiltonian in the momentum frame is given by:
\begin{equation}
\begin{aligned}
    \mathcal{H}_{kj}(s) &= \delta_{kj} (1-s) \epsilon_k \mathbb{I}_{\text{sys}} \\
    &\quad + \frac{1-s}{T+1} \tilde{\Pi}_{\text{init}} \\
    &\quad + \frac{s}{(T+1)^2} \mathcal{V}_k \mathcal{V}_j^\dagger \\
    &\quad + \frac{s}{T+1}\sum_{t=0}^T e^{-i(\omega_k - \omega_j)t} \mathcal{U}_t^\dagger H_{\text{target}} \mathcal{U}_t.
\end{aligned}
\end{equation}
    \bibliography{citations}
\end{document}